\documentclass{aa}
\usepackage{graphicx}
\usepackage{txfonts}
\usepackage{natbib}
\usepackage{color}
\usepackage[colorlinks, allcolors=blue]{hyperref}

\begin{document}

\title{Evolution Of Super-Sonic Downflows In A Sunspot}

\author{C. J. Nelson$^{1}$, S. Krishna Prasad$^{1,2}$, M. Mathioudakis$^{1}$}

\offprints{c.nelson@qub.ac.uk}
\institute{$^1$Astrophysics Research Centre (ARC), School of Mathematics and Physics, Queen’s University, Belfast, BT7 1NN, NI, UK.\\
$^2$Centre for mathematical Plasma Astrophysics (CmPA), KU Leuven, Celestijnenlaan 200B, 3001 Leuven, Belgium.}

\date{}

\abstract
{Super-sonic downflows have been observed in transition region spectra above numerous sunspots; however, little research has been conducted to date into how persistent these signatures are within sunspots on time-scales longer than a few hours.}
{Here, we aim to analyse the lead sunspot of AR $12526$ to infer the properties and evolution of super-sonic downflows occurring within it using high-spatial and spectral resolution data.}
{Sixteen large, dense raster scans sampled by the Interface Region Imaging Spectrograph are analysed. These rasters tracked the lead sunspot of AR $12526$ across the solar disc at discrete times between the $27$th March $2016$ and the $2$nd April $2016$, providing spectral profiles from the \ion{Si}{IV}, \ion{O}{IV}, \ion{Mg}{II}, and \ion{C}{II} lines. Additionally, one sit-and-stare observation acquired on the $1$st April $2016$ centred on the sunspot is studied in order to analyse the evolution of super-sonic downflows on shorter time-scales.}
{Super-sonic downflows are variable within this sunspot both in terms of spatial structuring and velocities. $13$ of the $16$ raster scans display some evidence of super-sonic downflows in the \ion{Si}{IV} $1394$ \AA\ line co-spatial to a sustained bright structure detected in the $1400$ \AA\ slit-jaw imaging channel, with a peak velocity of $112$ km s$^{-1}$ being recorded on the $29$th March $2016$. Evidence for super-sonic downflows in the \ion{O}{IV} $1401$ \AA\ line was found in $14$ of these rasters, with the spatial structuring in this line often differing from that inferred from the \ion{Si}{IV} $1394$ \AA\ line. Only one example of a super-sonic downflow was detected in the \ion{C}{II} $1335$ \AA\ line, with no downflows being found in the \ion{Mg}{II} $2796$ \AA\ lines at these locations. In the sit-and-stare observations, no dual flow is initially detected, however, a super-sonic downflow does develop after around $60$ minutes. This downflow accelerates from $73$ km s$^{-1}$  to close to $80$ km s$^{-1}$ in both the \ion{Si}{IV} $1394$ \AA\ and \ion{O}{IV} $1401$ \AA\ lines over the course of $20$ minutes before the end of the observation.}
{Super-sonic downflows were found in the \ion{Si}{IV} $1394$ \AA\ line in $13$ of the $16$ rasters studied here. The morphology of these downflows evolved over the course of both hours and days and was often different in the \ion{Si}{IV} $1394$ \AA\ and \ion{O}{IV} $1401$ \AA\ lines. These events were found co-spatial to a bright region in the core of the \ion{Si}{IV} $1394$ \AA\ line which appeared to form at the foot-points of coronal fan loops. Our results indicate that one raster is not enough to conclusively draw inferences about the properties of super-sonic downflows within a sunspot during its lifetime.}

\authorrunning{Nelson et al.}
\titlerunning{On-Set Of Super-Sonic Downflows In A Coronal Fan Loop}

\maketitle

\section{Introduction}
	\label{Introduction}

The solar atmosphere above sunspots is highly dynamic, with a range of transient events and oscillations being reported in the literature (for reviews see, for example, \citealt{Solanki03, Khomenko15}). These strong regions of vertical magnetic field (\citealt{Hale08}) act as an ideal conduit along which magneto-hydrodynamic (MHD) waves can propagate from the lower solar atmosphere into the upper solar atmosphere (examples include \citealt{Fludra01, Rouppe03, Jess12, Sych14}) potentially depositing their energy to heat the local corona. It is important to note, however, that not everything within umbrae is always moving upwards. It is now known that downflowing plasma can play an important role in sunspot dynamics, having been associated with strong shocks in both the chromosphere (\citealt{Henriques17, Nelson17}) and the transition region (as discussed by, e.g., \citealt{Dere82, Brynildsen01, Brynildsen04, Kleint14}). 

\begin{figure*}
\includegraphics[width=0.98\textwidth]{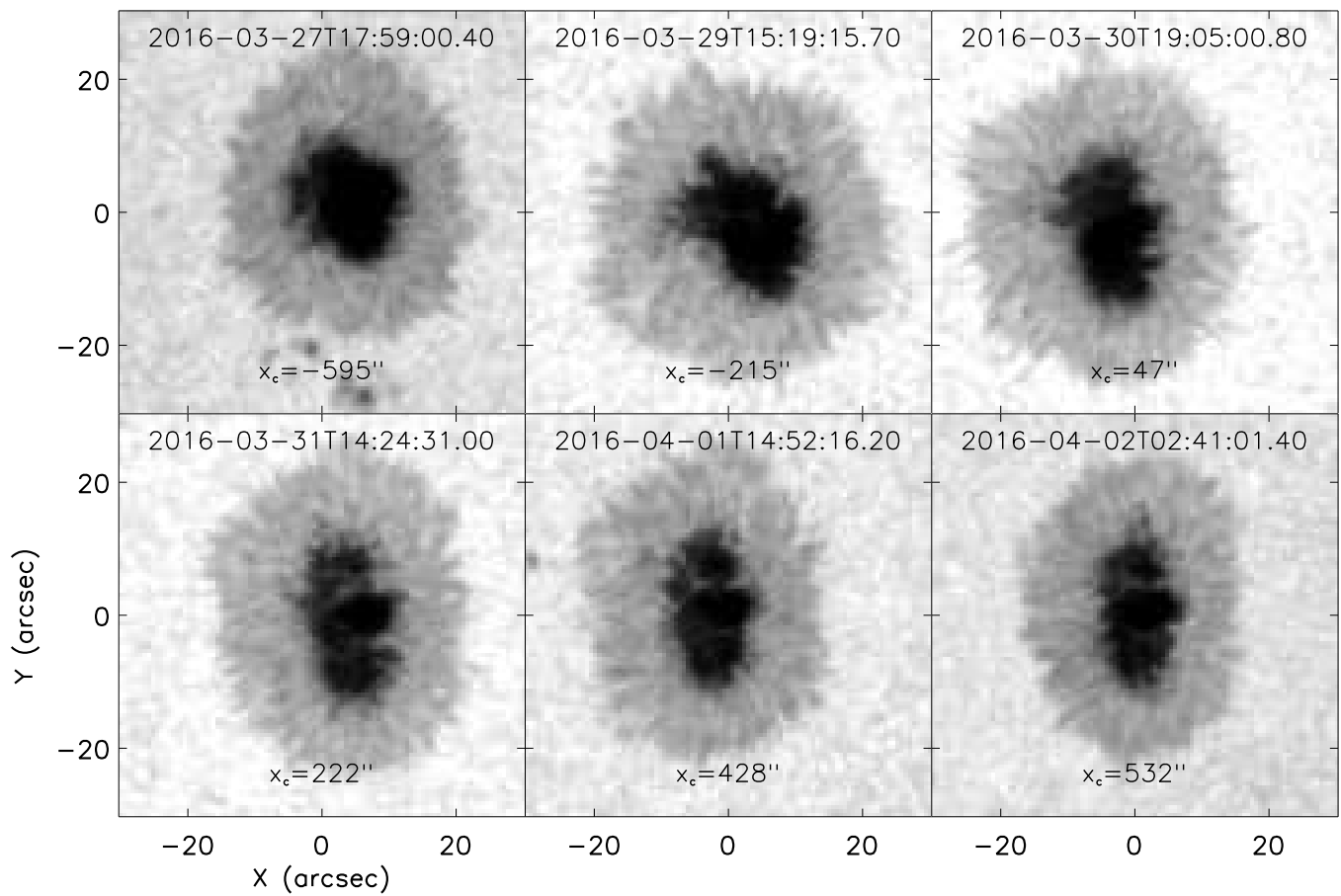}
\caption{Context imaging from the SDO/HMI continuum channel of the sunspot within AR $12526$ over the course of the seven days analysed here. Initially, no lightbridge was evident, however, by the $31$st March $2016$ one large lightbridge had developed in the northern region of the sunspot. A second lightbridge was also present in the southern portion of the sunspot, however, this was smaller and less obvious than the northern lightbridge. The times and co-ordinates over-laid on each panel indicate the start time and centre position in the $x-$direction of the corresponding IRIS raster.}
\label{Overview_rast}
\end{figure*}

Dynamic shocks in the solar chromosphere such as umbral flashes have been linked to upflowing plasma for decades (see, for example, \citealt{Beckers69, Wittmann69, Rouppe03}). However, recent non-LTE modelling of chromospheric lines using inversion codes such as NICOLE (\citealt{Socas15}) has indicated that such events could be formed by a blend of an upwardly propagating wave on a strongly downflowing background (as discussed by \citealt{Henriques17, Bose19}). Interestingly, such a model could also be invoked to simultaneously explain other events in the chromospheres of sunspots including Small-Scale Umbral Brightenings (\citealt{Nelson17}) and short dynamic fibrils (\citealt{Rouppe13}). Downflows have also been observed above rapidly evolving pores using the \ion{He}{I} $10830$ \AA\ line (\citealt{Lagg07}).

Super-sonic downflows in the transition region above sunspots have been of interest to the community ever since their discovery (see, for example, \citealt{Dere82, Nicolas82, Kjeldseth88, Brynildsen01, Brynildsen04, Tian14, Kleint14, Samanta18}). These `dual flows', which manifest as strongly red-shifted secondary emission peaks in transition region spectra such as \ion{Si}{IV} and \ion{O}{IV}, form in localised regions of both the umbra and penumbra (as shown by the recent statistical analysis of \citealt{Samanta18}). It has been suggested that these super-sonic downflows are associated with lightbridges in the lower solar atmosphere, however, the details of such a relationship are currently unknown (\citealt{Nicolas82}). The extensive availability of high-spectral, temporal, and spatial resolution data since the launch of the Interface Region Imaging Spectrograph (IRIS; \citealt{dePontieu14}) has allowed improved understanding of these events over recent years (for a recent review see \citealt{Tian18}).

It is now thought possible that at least two different types of dual flows may exist in the transition region. Bursty, short-lived (of the order seconds) downflows with velocities of up to $200$ km s$^{-1}$ have been detected and associated with impulsive coronal rain (see, for example, \citealt{Kleint14}). Typically, these events appear to have signatures in a wide complement of IRIS diagnostics, including the \ion{Mg}{II}, \ion{C}{II}, and \ion{Si}{IV} lines. Of more relevance here, potentially, are downflows which appear to have lifetimes of the order minutes to hours and typical velocities of around $100$ km s$^{-1}$ (as discussed by, e.g., \citealt{Straus15, Chitta16}). These events, which are potentially linked to siphon flows, are generally observed in the \ion{Si}{IV} and \ion{O}{IV} lines with only occasional signatures in the \ion{Mg}{II} and \ion{C}{II} lines (\citealt{Samanta18}). Although these events have sometimes been observed to be stable over time-scales of the order hours, it is likely that they will evolve significantly over the course of several days (\citealt{Kjeldseth88}). Interestingly, the inferred downflow velocity from the \ion{O}{IV} lines has sometimes been found to be around $10$ km s$^{-1}$ lower than the velocity calculated from the \ion{Si}{IV} lines potentially implying the presence of multi-thermal, multi-threaded loops (\citealt{Dere82, Nicolas82, Chitta16}). It should be noted, however, that such a disparity in the downflow speeds measured from these spectral lines is not always present (\citealt{Samanta18}).

\begin{figure*}
\includegraphics[width=0.98\textwidth]{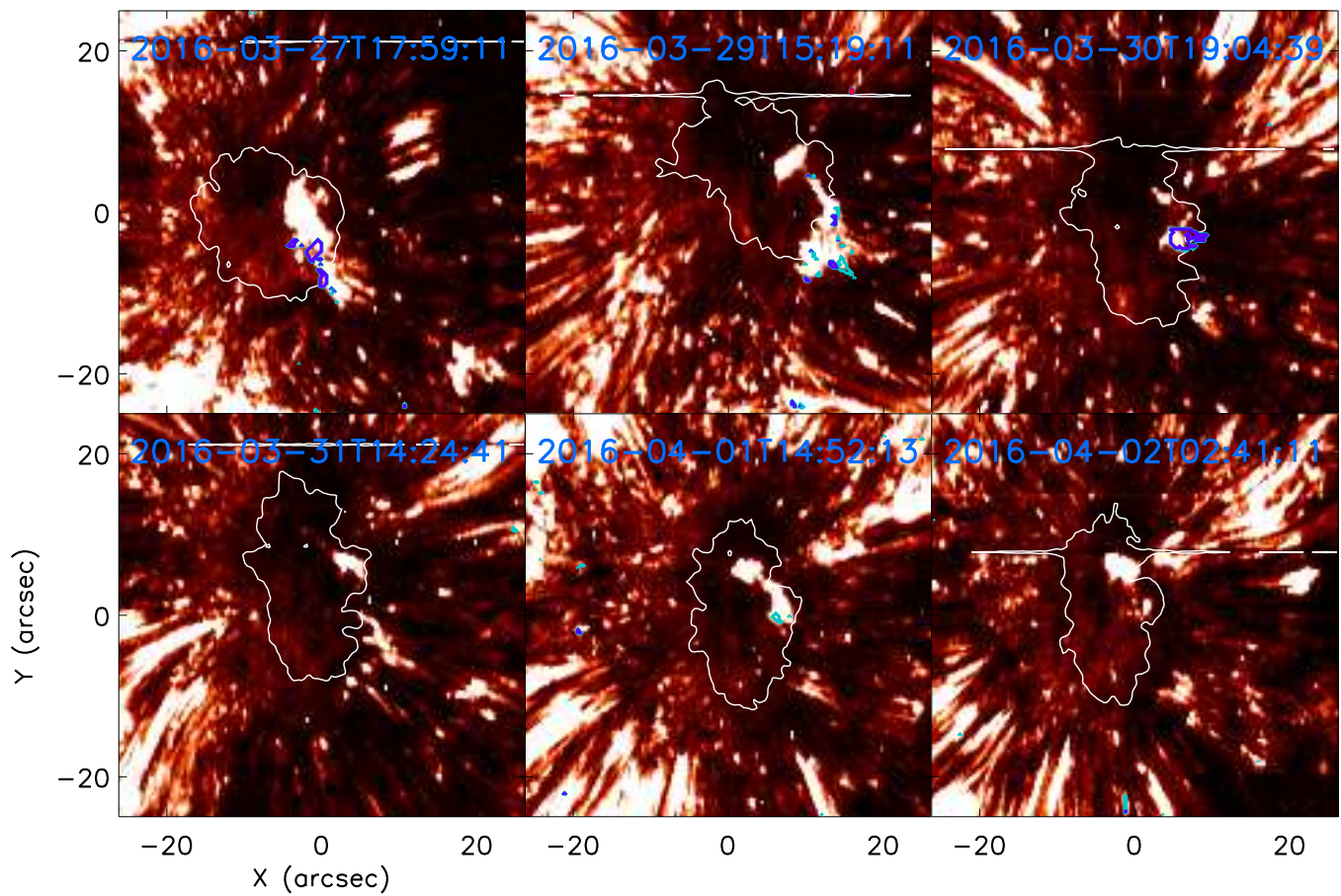}
\caption{The intensity at the rest wavelength of the \ion{Si}{IV} $1394$ \AA\ line in the sunspot from the rasters corresponding to the SDO/HMI continuum panels in Fig.~\ref{Overview_rast}. The over-laid coloured contours highlight regions where double Gaussian fitting indicated the presence of super-sonic downflows (aqua and blue correspond to $50$ km s$^{-1}$ and $75$ km s$^{-1}$, respectively). All super-sonic downflows within the sunspot are co-spatial to a sustained region of increased intensity in the line core. Again, pointing coordinates are not relative and are only provided for an indication of scale. The white contours plot the outline of the sunspot as inferred from the average intensity in the continuum between the two \ion{Mg}{II} lines. It should be noted that not all rasters covered the entirety of the FOV plotted in Fig.~\ref{Overview_rast} meaning the FOV plotted here is close to but not perfectly aligned with that figure.}
\label{Spect_rast}
\end{figure*}

In this article, we analyse the evolution of super-sonic downflows within the lead sunspot of AR $12526$ over the course of one week using high-resolution IRIS data. Of specific interest here is understanding how persistent such downflows are within the sunspot, over both short and long time-scales. Our work is set out as follows: In Sect.~\ref{Observations} we introduce the satellite data studied here; In Sect.~\ref{Results} we present our results, including analysis of both raster data and sit-and-stare observations; In Sect.~\ref{Conclusions} we present a discussion and our conclusions; Finally, in Sect.~\ref{Summary} we provide a brief summary.

\section{Observations}
	\label{Observations}

Initially, sixteen $400$-step dense ($0.35$\arcsec\ step-size) rasters covering AR $12526$ at discrete times between $27$th March $2016$ and $2$nd April $2016$ were studied. For each of these datasets the slit length was $175$\arcsec, the exposure time was $8$ s (with a step cadence of $9.2$ s), and the spectral samplings were $0.025$ \AA\ and $0.05$ \AA\ for the FUV and NUV channels, respectively. During the rasters, the Slit-Jaw Imager (SJI) sampled each of the $1330$ \AA, $1400$ \AA, $2796$ \AA, and $2832$ \AA\ channels sequentially, providing $100$ images per raster for each filter. Each SJI image sampled a $167$\arcsec$\times175$\arcsec\ field-of-view (FOV) which tracked the IRIS slit, giving a total FOV of $307$\arcsec$\times175$\arcsec. The cadence of each of the slit-jaw channels was approximately $37$ s and the pixel scale was $0.33$\arcsec. The basic information for each of these datasets is included for reference in the first five columns of Table~\ref{Tab1}. The OBS ID for each raster was: $3600108078$.

\begin{figure*}
\includegraphics[width=0.98\textwidth]{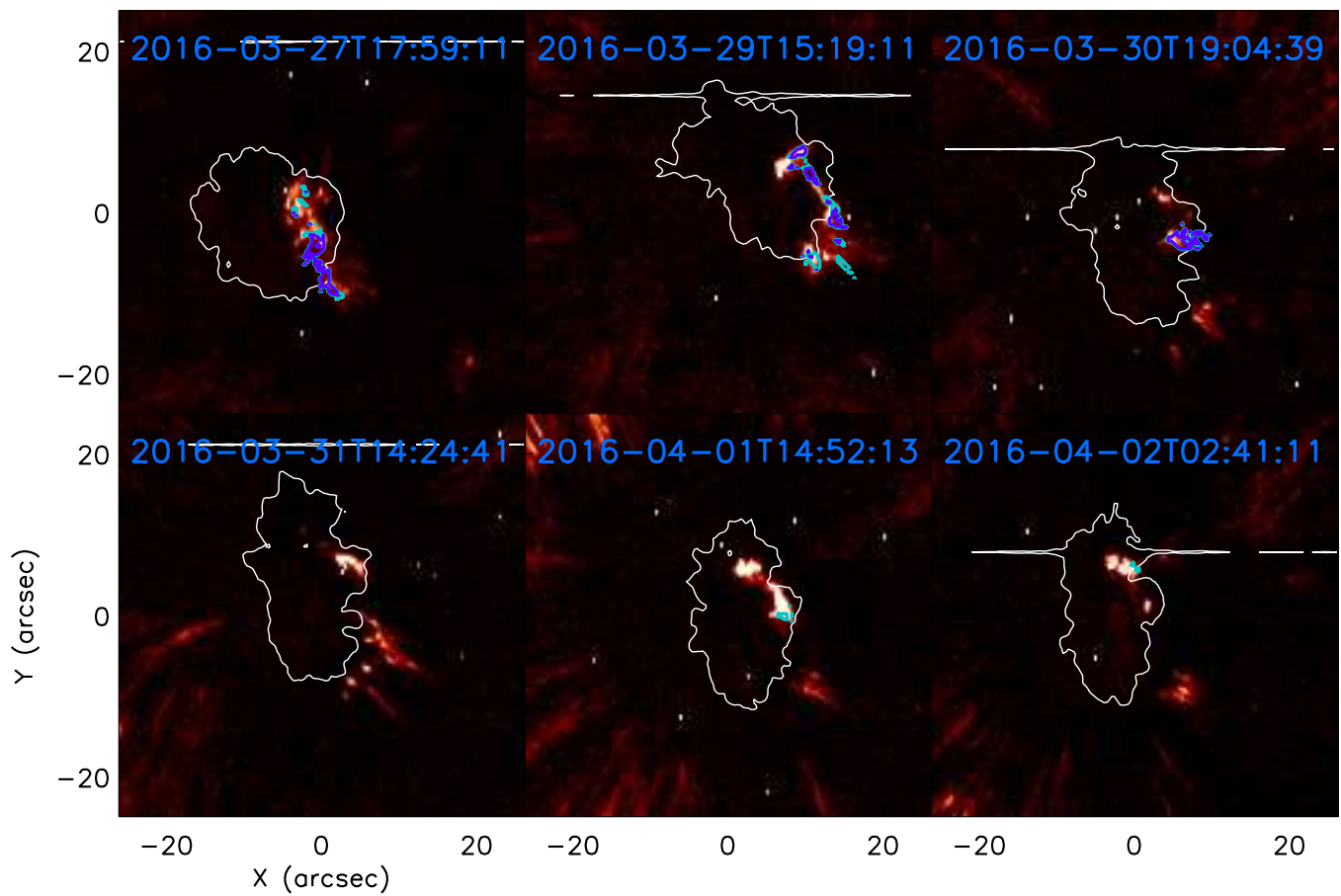}
\caption{Same as Fig.~\ref{Spect_rast} but for the \ion{O}{IV} $1401$ \AA\ line. Note that the spatial coverage of downflows in the top left and top middle panels is larger than for the respective \ion{Si}{IV} $1394$ \AA\ plots. Additionally, the bottom right panel displays evidence of a super-sonic downflow in the \ion{O}{IV} $1401$ \AA\ line when no downflow was detected in the \ion{Si}{IV} $1394$ \AA\ line for this raster.}
\label{Spect_rast_O}
\end{figure*}

To supplement the raster data, we also analysed one sit-and-stare sequence sampled by IRIS between $16$:$23$:$28$ UT and $17$:$44$:$22$ UT on the $1$st April $2016$. During this time, the slit passed north-south through the umbra of the lead sunspot within AR $12526$ at coordinates of $x_\mathrm{c}$=$480$\arcsec, $y_\mathrm{c}$=$34$\arcsec\ (centre of the slit). These data had a $15$ s exposure time (giving a cadence of $16.7$ s) returning a total of $290$ spectra. The slit length was $119$\arcsec\ and the spectral sampling in both the NUV and FUV channels was approximately $0.025$ \AA. The SJI sampled a $119$\arcsec$\times119$\arcsec\ FOV with the $1400$ \AA\ and $2796$ \AA\ channels sequentially, meaning each filter had a cadence of approximately $33$ s. The pixel scale of these SJI data was around $0.17$\arcsec. The OBS ID for this sit-and-stare sequence was $3680261403$. Wavelength calibration for the FUV and NUV channels was confirmed for all IRIS datasets through analysis of the neutral \ion{O}{I} $1355.6$ \AA\ and \ion{Ni}{I} $2799.5$ \AA\ lines.

Finally, we studied data from the Solar Dynamics Observatory's Helioseismic and Magnetic Imager (SDO/HMI; \citealt{Scherrer12}) and Atmospheric Imaging Assembly (SDO/AIA; \citealt{Lemen12}) to give photospheric and coronal context to our analysis. Single line-of-sight (LOS) magnetic field maps and continuum intensity images from the SDO/HMI instrument co-temporal to the start of each raster and sit-and-stare sequence were downloaded and reduced, with a post-reduction pixel scale of $0.6$\arcsec\ and a cadence of $45$ s. Additonally, SDO/AIA $304$ \AA, $171$ \AA, and $193$ \AA\ filter images were downloaded for the time-period corresponding to the entire sit-and-stare sequence. with post-reduction pixel scales and cadences of $0.6$\arcsec\ and $12$ s, respectively. A single SDO/AIA $1600$ \AA\ image co-temporal to the start of the sit-and-stare sequence was also downloaded for alignment purposes. Initial alignment between all instruments was accomplished to the leading order using header information before more accurate alignments were completed by visually matching features in the SDO/HMI continuum, SDO/AIA $1600$ \AA, and IRIS SJI \ion{Mg}{II} $2796$ \AA\ channels.

\section{Results}
	\label{Results}

\subsection{Raster Scans}

We begin our analysis by providing a brief overview of the general properties of the sunspot studied here through time. In Fig.~\ref{Overview_rast}, we plot the sunspot as sampled by the SDO/HMI continuum filter at the start of six of the raster scans detailed in Table~\ref{Tab1}, one for each day in which a raster exits. On the $27$th March $2016$, no lightbridge is evident within the umbra of what appears to be a relatively circular sunspot. Over the course of the next three days, however, the sunspot becomes more elongated in the north-south direction before two distinct lightbridges develop on the $31$st March $2016$. Both lightbridges have an east-west orientation with one forming in the north and one forming in the south, essentially splitting the umbra into three equal portions. The northern lightbridge has a higher contrast than the southern lightbridge making it more evident in the SDO/HMI images plotted here. Both lightbridges are present up to the time of the final raster at $02$:$41$ UT on the $2$ April $2016$.

\begin{table*}
\begin{center}
\begin{tabular}{| c | c | c | c | c | c | c | c | c | c | c | c |}
\hline
\bf{Date} & \bf{Start (UT)} & \bf{End (UT)} & \bf{xc\arcsec} & \bf{yc\arcsec} & Pix$_\mathrm{x}$ & Pix$_\mathrm{y}$ & \ion{Si}{IV} $1394$ \AA\ & \ion{O}{IV} $1401$ \AA\ & \ion{C}{II} $1335$ \AA\  & $n_\mathrm{p,Si}$ & $n_\mathrm{p,O}$ \\ \hline
2016-03-27 & 17:59:11 & 19:00:25 & -595 & 18 & 313 & 324 & 98 km s$^{-1}$ & 99 km s$^{-1}$ & No & 62 & 127 \\ \hline
2016-03-29 & 01:29:12 & 02:30:25 & -323 & 29 & 302 & 307 & 72 km s$^{-1}$ & 72 km s$^{-1}$ & No & 25 & 55 \\ \hline
2016-03-29 & 06:13:11 & 07:14:25 & -304 & 16 & 384 & 369 & 107 km s$^{-1}$ & 108 km s$^{-1}$ & No & 73 & 122 \\ \hline
2016-03-29 & 09:29:11 & 10:30:25 & -268 & 16 & 375 & 361 & 112 km s$^{-1}$ & N/A & No & 11 & 56 \\ \hline
2016-03-29 & 15:19:11 & 16:20:25 & -215 & 26 & 319 & 354 & 94 km s$^{-1}$ & N/A & No & 61 & 136 \\ \hline
2016-03-30 & 01:09:11 & 02:10:25 & -126 & 19 & 358 & 363 & 89 km s$^{-1}$ & 87 km s$^{-1}$ & No & 16 & 65 \\ \hline
2016-03-30 & 08:19:09 & 09:20:23 & -58 & 28 & 346 & 354 & 94 km s$^{-1}$ & 86 km s$^{-1}$ & No & 2 & 6 \\ \hline
2016-03-30 & 19:04:39 & 20:05:53 & 47 & 25 & 356 & 343 & 85 km s$^{-1}$ & 88 km s$^{-1}$ & 100 km s$^{-1}$ & 75 & 60 \\ \hline
2016-03-31 & 01:29:09 & 02:30:23 & 109 & 27 & 341 & 331 & 76 km s$^{-1}$ & 85 km s$^{-1}$ & No & 64 & 57 \\ \hline
2016-03-31 & 08:34:11 & 09:35:25 & 171 & 25 & 340 & 338 & 87 km s$^{-1}$ & 85 km s$^{-1}$ & No & 24 & 6 \\ \hline
2016-03-31 & 14:24:41 & 15:25:55 & 222 & 29 & N/A & N/A & N/A & N/A & N/A & 0 & 0 \\ \hline
2016-03-31 & 19:22:11 & 20:23:25 & 266 & 22 & N/A & N/A & N/A & N/A & N/A & 0 & 0 \\ \hline
2016-04-01 & 05:50:11 & 06:51:25 & 355 & 40 & 297 & 330 & 85 km s$^{-1}$ & 69 km s$^{-1}$ & No & 18 & 12 \\ \hline
2016-04-01 & 14:52:13 & 15:53:27 & 428 & 32 & 294 & 341 & 71 km s$^{-1}$ & N/A & No & 9 & 14 \\ \hline
2016-04-01 & 21:20:11 & 22:21:25 & 485 & 9 & 353 & 327 & 57 km s$^{-1}$ & 60 km s$^{-1}$ & No & 7 & 20 \\ \hline
2016-04-02 & 02:41:11 & 03:42:25 & 532 & 1 & 384 & 304 & N/A & 61 km s$^{-1}$ & N/A & 0 & 5 \\ \hline
\end{tabular}
\end{center}

\caption{Details of the $16$ raster scans analysed here including: Date of raster; Start time of raster; End time of raster; $x$-coordinate of the centre of the raster; $y$-coordinate of the centre of the raster; $x$-pixel where the largest super-sonic downflow was inferred from the \ion{Si}{IV} $1394$ \AA\ line; $y$-pixel where the largest super-sonic downflow was inferred from the \ion{Si}{IV} $1394$ \AA\ line; Peak downflow velocity calculated from double Gaussian fitting of the \ion{Si}{IV} $1394$ \AA\ line; Downflow velocity in the \ion{O}{IV} $1401$ \AA\ line co-spatial to the peak \ion{Si}{IV} $1394$ \AA\ velocity (except bottom row where the peak \ion{O}{IV} $1401$ \AA\ velocity is reported); Downflow velocity in the \ion{C}{II} $1335$ \AA\ channel co-spatial to the peak \ion{Si}{IV} $1394$ \AA\ velocity; Number of pixels displaying downflows in the \ion{Si}{IV} $1394$ \AA\ line; Number of pixels displaying downflows in the \ion{O}{IV} $1401$ \AA\ line.}
\label{Tab1}
\end{table*}

In order to investigate whether super-sonic downflows were present in the transition region of the sunspot during these scans, we perfomed double Gaussian fitting to the \ion{Si}{IV} $1394$ \AA\ and \ion{O}{IV} $1401$ \AA\ spectra (using $gauss\_fit.pro$) on each pixel around the sunspot between velocities of $-45$ km s$^{-1}$ and $+175$ km s$^{-1}$. For the \ion{Si}{IV} $1394$ \AA\ line, we considered that a super-sonic downflow had been detected when the peak intensity of the line was larger than $50$ DN in this spectral range and when a Gaussian was returned between $50$ km s$^{-1}$ and $150$ km s$^{-1}$ with a peak intensity larger than $30$ DN and a width (in Doppler units) of larger than $10$ km s$^{-1}$. For the \ion{O}{IV} $1401$ \AA\ line, the required initial thresholding intensity of the line was $30$ DN, the maximum returned velocity was $100$ km s$^{-1}$ (larger values were searched for manually but were only detected once and, therefore, this lower peak velocity was implemented to minimise the effects of noise in the fitting), and the secondary emission peak was required to have an intensity of over $20$ DN. The background transition region intensity of the FOV was estimated through the construction of two-dimensional maps of the \ion{Si}{IV} $1394$ \AA\ and \ion{O}{IV} $1401$ \AA\ line intensities at their rest wavelengths. 

In Fig.~\ref{Spect_rast}, we plot six examples of the \ion{Si}{IV} $1394$ \AA\ core intensity maps, where each panel corresponds to the respective panel in Fig.~\ref{Overview_rast}. An extended bright region is clear in the sunspot (the outline of the umbra as inferred from the \ion{Mg}{II} continuum between the two spectral lines is over-laid in white) stretching above both the umbra and penumbra throughout the time-period analysed here. The coloured contours outline regions where downflows were detected with aqua indicating a velocity of over $50$ km s$^{-1}$ and blue indicating a velocity of over $75$ km s$^{-1}$. Applying the typical formation temperatures of the \ion{Si}{IV} $1394$ \AA\ line ($10^{4.8}-10^{5}$ K) to the formula $C_\mathrm{s}$=$152$ T$^{0.5}$ m s$^{-1}$ (\citealt{Priest84}) allows us to infer that downflow velocities over $50$ km s$^{-1}$ will be super-sonic. It should be noted that the FOV in Fig.~\ref{Spect_rast} and the FOV in Fig.~\ref{Overview_rast} are not perfectly aligned due to the different rasters sampling slightly different regions around the sunspot with some rasters only partially covering the penumbra. The choice was made to have this slight off-set in the FOVs to better emphasise the evolution of the entire spot in Fig.~\ref{Overview_rast}.

In Fig.~\ref{Spect_rast_O}, we reproduce Fig.~\ref{Spect_rast} but with the background intensity and velocity contours calculated from the \ion{O}{IV} $1401$ \AA\ line. It is immediately evident that super-sonic downflows in the \ion{O}{IV} $1401$ \AA\ line covered a larger spatial extent than super-sonic downflows in the \ion{Si}{IV} $1394$ \AA\ line in the top left and top middle panels, with downflows occurring along almost the entirety of the transition region brightening. To quantify this, we calculated the number of pixels within which super-sonic downflows over $50$ km s$^{-1}$ were detected for both the \ion{Si}{IV} $1394$ \AA\ and \ion{O}{IV} $1401$ \AA\ lines (the criteria defined in the second paragraph of this section). In the top left panel, super-sonic downflows were detected in $62$ and $127$ pixels respectively. Within the top middle panel the corresponding number of pixels for the two lines was $61$ and $136$. Although such a large disparity in the number of pixels displaying super-sonic downflows in the \ion{Si}{IV} $1394$ \AA\ and \ion{O}{IV} $1401$ \AA\ lines is not apparent in each raster studied here (see final two columns of Table~\ref{Tab1}), this result does confirm that the spatial structuring of these dual flows can be different in different spectral (i.e., thermal) windows. Additionally, the bottom right panel displays evidence of a super-sonic downflow in the \ion{O}{IV} $1401$ \AA\ line which is not present in the \ion{Si}{IV} $1394$ \AA\ line, further highlighting the difference in spatial structuring between these wavelengths.

\begin{figure}
\includegraphics[width=0.99\columnwidth]{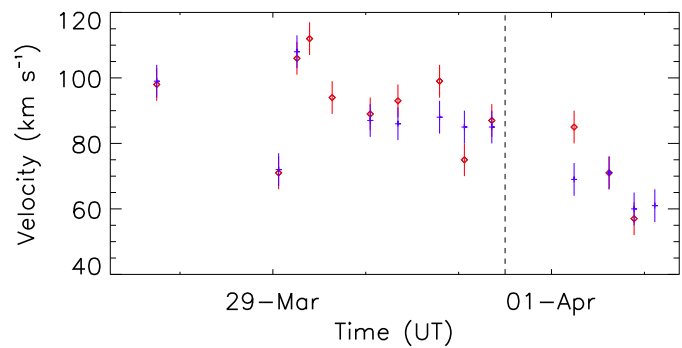}
\caption{The peak super-sonic downflow velocities inferred from the \ion{Si}{IV} $1394$ \AA\ (red squares) for each raster. The blue crosses indicate the co-spatial \ion{O}{IV} $1401$ \AA\ downflow veloc-ity. The error bars highlight the spectral sampling and the dashed vertical line indicates the approximate time of the lightbridge formation.}
\label{Int_time}
\end{figure}

Of the $16$ raster scans analysed here, $13$ ($14$) displayed evidence of downflows with velocities larger than $50$ km s$^{-1}$ in the \ion{Si}{IV} $1394$ \AA\ (\ion{O}{IV} $1401$ \AA) line. All detected super-sonic downflows formed co-spatial to the intense structure at the centre of the sunspot easily evident in the \ion{Si}{IV} $1394$ \AA\ line core (see Fig.~\ref{Spect_rast}), with no dual flows being detected in the sunspot away from this transition region brightening. The evolution of the super-sonic downflows through time exhibited no clear pattern, with the spatial positioning along the brightening and the sizes of the detected downflows changing considerably from raster to raster. To display this, we plot the peak downflow velocity in the \ion{Si}{IV} $1394$ \AA\ line (red squares) and the co-spatial, co-temporal downflow velocities from the \ion{O}{IV} $1401$ \AA\ line (blue crosses) in Fig.~\ref{Int_time}. We note that super-sonic downflows were not detected in the \ion{O}{IV} $1401$ \AA\ line at the sites of the peak \ion{Si}{IV} $1394$ \AA\ downflow velocities in three rasters (see Table~\ref{Tab1} for detailed information). The error bars indicate the spectral sampling of the data and the dashed vertical line indicates the approximate formation time of the lightbridges.

\begin{figure*}
\includegraphics[width=0.98\textwidth]{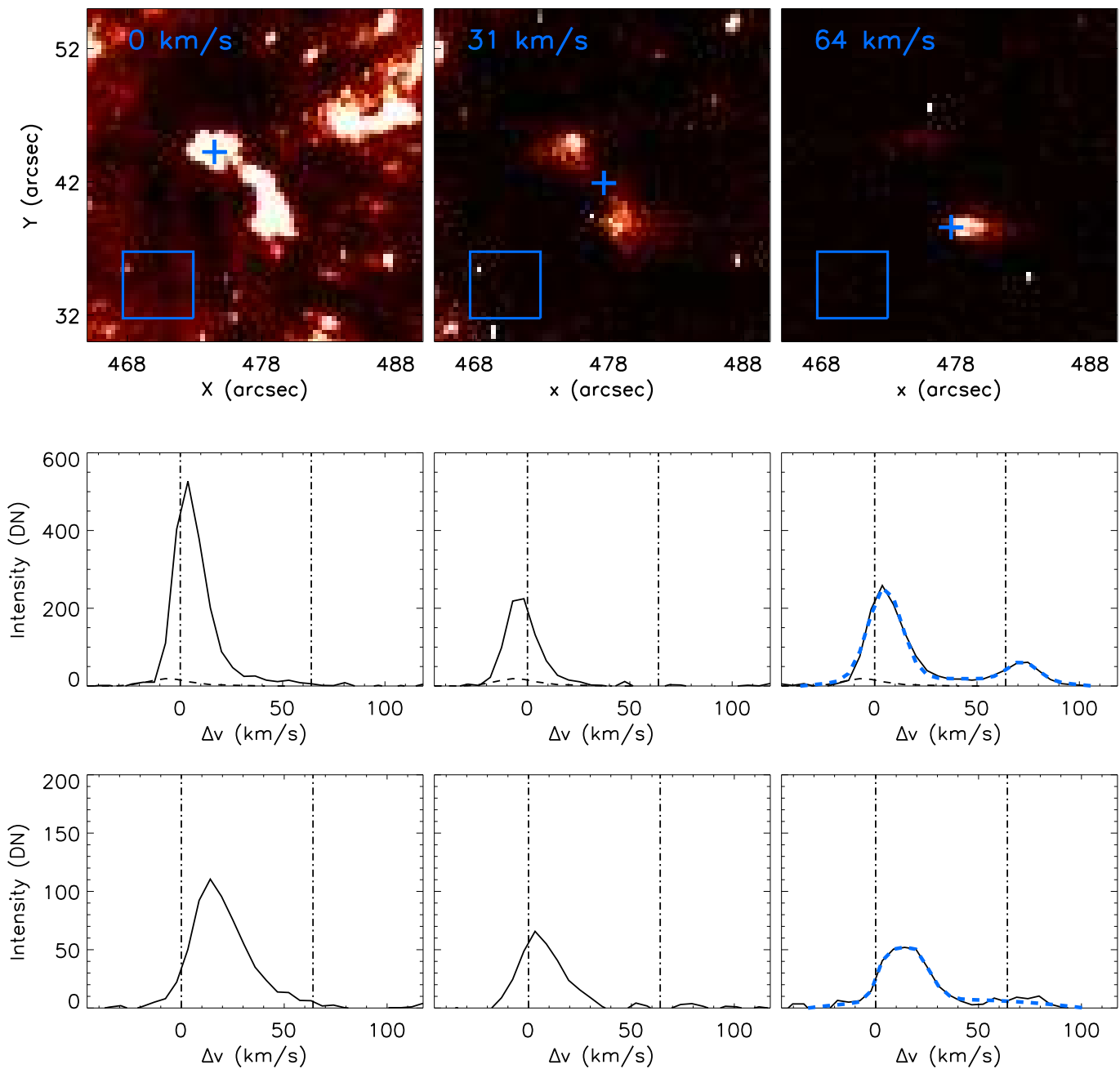}
\caption{(Top row) \ion{Si}{IV} $1394$ \AA\ intensity maps at three different positions within the line profile for the raster beginning at $14$:$52$ on the $1$st April $2016$. The LOS velocities corresponding to the specific line positions are indicated in the top left of each panel. (middle row) \ion{Si}{IV} $1394$ \AA\ spectra for the spatial positions indicated by the crosses in the top row. (Bottom row) Same as the middle row but for the \ion{O}{IV} $1401$ \AA\ line. Average spectra from the sunspot umbra, calculated from the boxes in the bottom left of the images in the top row, are over-plotted as dashed lines. The vertical lines indicate the rest velocity of the lines and a LOS velocity of $64$ km s$^{-1}$ (matching the line position plotted in the top right panel). The dashed blue lines on the right hand panels of the bottom two rows plot the double Gaussian fits for this pixel. As previously reported in Table~\ref{Tab1}, no clear evidence of a secondary emission peak is found in the \ion{O}{IV} $1401$ \AA\ line at this location.}
\label{Loop_Rast}
\end{figure*}

Each of the ten datasets sampled prior to the appearance of the lightbridges ($14$:$24$ on the $31$st March $2016$) displayed some evidence of super-sonic downflows in the \ion{Si}{IV} $1394$ \AA\ line co-spatial to the sunspot transition region brightening (see Table~\ref{Tab1}). Downflow velocities of over $70$ km s$^{-1}$ were detected in all of these datasets, with velocities of over $100$ km s$^{-1}$ even being evident in two rasters (the consecutive scans starting at $06$:$13$ UT and $09$:$29$ UT on the $29$th March $2016$). As the sunspot rotated from $x_\mathrm{c}$=$-595$\arcsec\ to $x_\mathrm{c}$=$171$\arcsec\ during this time we comparing the projected and deprojected downflow velocities to confirm that the $\mu$-angle of the observation did not affect our results. We found only small change in the estimated downflow velocity of around $5$ km s$^{-1}$ (comparable to the spectral sampling). The spatial structuring of the detected dual flows in these rasters was extremely variable with large areas of downflows ($65<$ pixels) being detected in the \ion{Si}{IV} $1394$ \AA\ line in five datasets and only small ($<25$ pixels) areas of downflows being evident in the remaining five raster scans. 

\begin{figure*}
\includegraphics[width=0.98\textwidth]{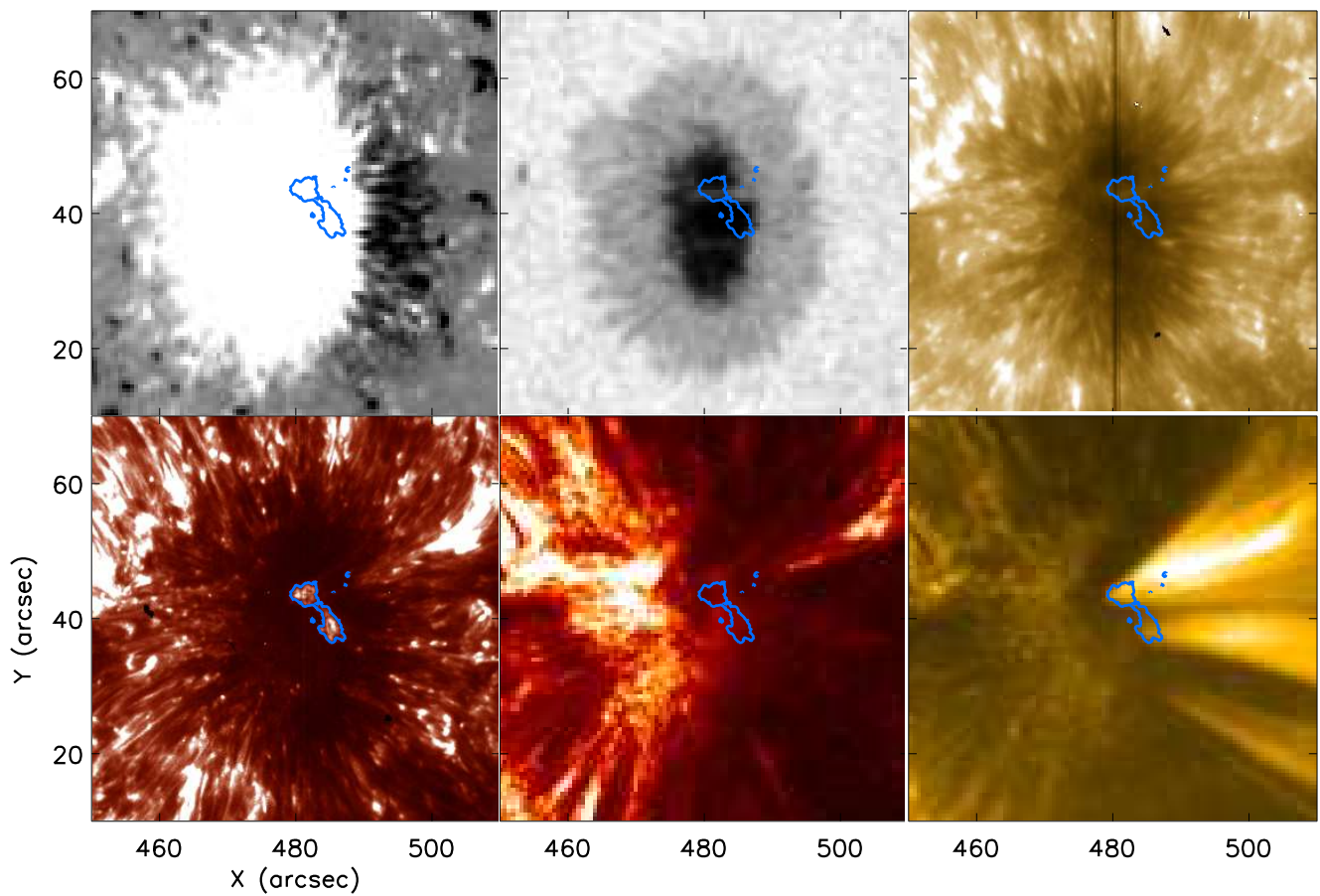}
\caption{The sunspot within AR $12526$ at approximately $16$:$23$:$28$ UT on the $1$st April $2016$. The photospheric LOS magnetic field and continuum intensity (top left and middle panels) sampled by the SDO/HMI instrument display a large positive polarity sunspot. The solar atmosphere above the sunspot was sampled using the IRIS \ion{Mg}{II} $2796$ \AA\ (top right) and \ion{Si}{IV} $1400$ \AA\ (bottom left) SJI filters, as well as the SDO/AIA $304$ \AA\ (bottom middle) and $171$ \AA\ (bottom right) channels. The IRIS slit position during the sit-and-stare sequence passes directly through the sunspot and is most easily evident in the $2796$ \AA\ image. The blue contours outline the bright feature in the IRIS $1400$ \AA\ channel within which all super-sonic downflows are detected in this sunspot.}
\label{Overview_SJI}
\end{figure*}

Following the appearance of the lightbridges, however, super-sonic downflows became spatially much less prevalent within the sunspot (see the bottom row of Fig.~\ref{Spect_rast}). Indeed, all three datasets for which no downflows of over $50$ km s$^{-1}$ were detected in the \ion{Si}{IV} $1394$ \AA\ line occurred following the appearance of the lightbridges (the final six rows of Table~\ref{Tab1}). The super-sonic downflows in the further three rasters were extremely localised with spatial extents of only a few pixels ($18$, $9$, and $7$ pixels each in the \ion{Si}{IV} $1394$ \AA\ line), while only one displayed evidence of a super-sonic downflow with a velocity of over $75$ km s$^{-1}$. Overall, the average peak downflow velocity in the \ion{Si}{IV} $1394$ \AA\ line prior to the formation of the lightbridges is $91.4$ km s$^{-1}$ compared to $71$ km s$^{-1}$ (calculated from the three rasters which displayed super-sonic downflows) after the formation of the lightbridges. We cannot confidently comment on whether the change in the super-sonic velocities apparent here is caused by the formation of the lightbridges or not due to the small number of scans collected both before and after this time, however, we can conclusively say that the presence of lightbridges in a sunspot is not a required condition for the occurrence of super-sonic downflows in the transition region (hypothesised by \citealt{Nicolas82}).

To further investigate these super-sonic downflows, we examined the spectra of the \ion{O}{IV} $1401$ \AA, \ion{C}{II} $1335$ \AA, and \ion{Mg}{II} $2796$ \AA\ lines co-spatial to the peak downflow velocity detected in the \ion{Si}{IV} $1394$ \AA\ line for each raster (pixel locations included in Table~\ref{Tab1}). Super-sonic downflows were present at these locations in the \ion{O}{IV} $1401$ \AA\ line for $10$ of the $13$ datasets, with the velocities (calculated by fitting double Gaussians at these locations) returned being comparable to the velocities inferred from the \ion{Si}{IV} $1394$ \AA\ line (agreeing with the results of \citealt{Samanta18}). The core intensity of the \ion{O}{IV} $1401$ \AA\ line for two of the datasets in which no super-sonic downflows were detected were extremely low meaning any dual peaks in the line wing would likely be below the noise level. Additionally, one example of a super-sonic downflow, with a peak velocity of $61$ km s$^{-1}$, was detected in the \ion{O}{IV} $1401$ \AA\ line in a raster when no \ion{Si}{IV} $1394$ \AA\ downflow was apparent (bottom row of Table~\ref{Tab1}). One example of a downflow was inferred from the \ion{C}{II} $1335$ \AA\ line, with the velocity of $100$ km s$^{-1}$ being $15$ km s$^{-1}$ faster than the local \ion{Si}{IV} $1394$ \AA\ downflow (see Table~\ref{Tab1}). From this one observation we are unable to comment on the generality or significance of this result. No signature of super-sonic downflows was detected in the \ion{Mg}{II} $2796$ \AA\ lines co-spatial to these dual flows.

\begin{figure*}
\includegraphics[width=0.98\textwidth]{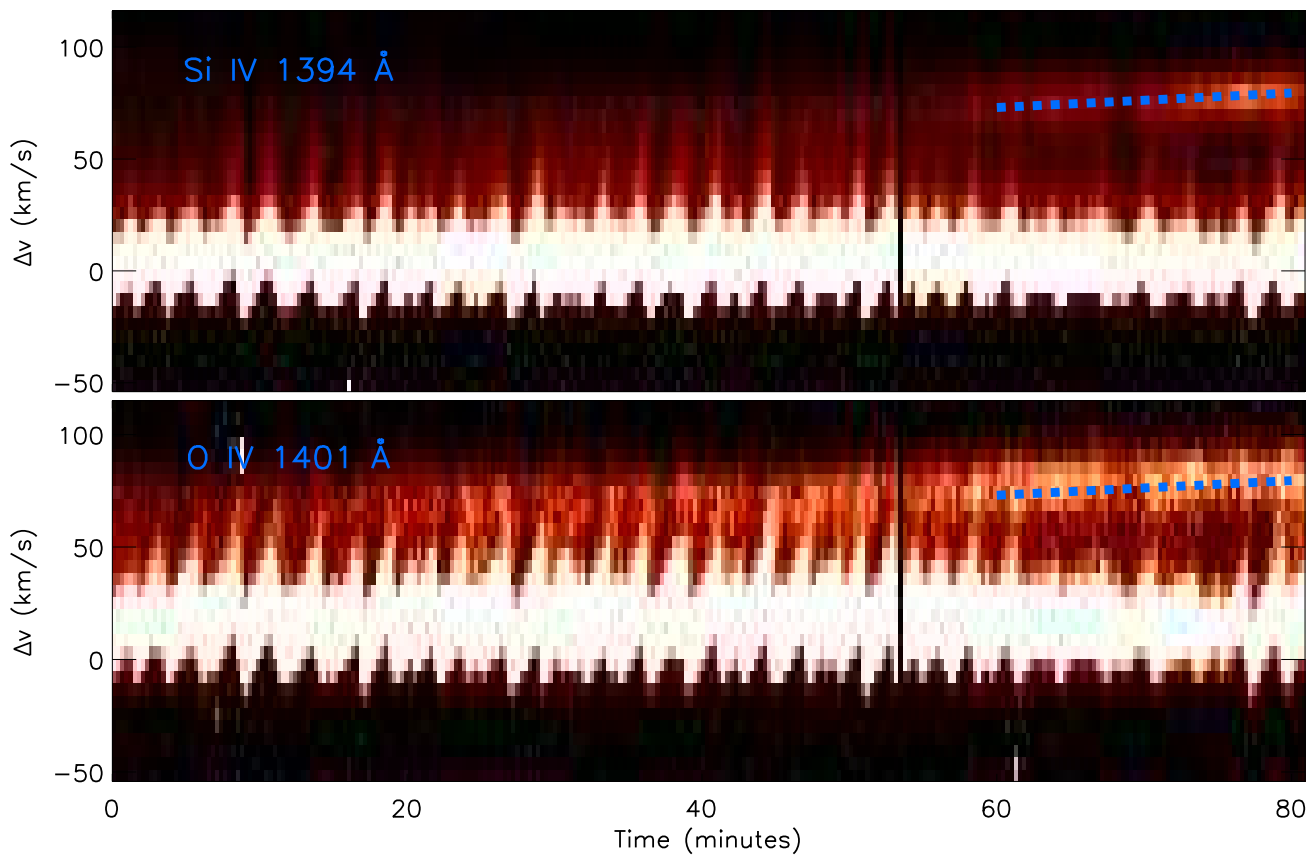}
\caption{(Top row) Spectral-time plot for the \ion{Si}{IV} $1394$ \AA\ line for a pixel within the transition region brightening. The typical saw-tooth shock pattern is clearly evident close to the line core. The super-sonic downflow can be detected from around 50 minutes until the end of the time-series. During this time, the flow accelerates at approximately $5$ m s$^{-2}$ from around $73$ km s$^{-1}$ to close to $80$ km s$^{-1}$, as indicated by the dashed blue line. (Bottom row) Same as the top row except for the \ion{O}{IV} $1401$ \AA\ line.}
\label{Sp_time}
\end{figure*}

Finally, we provide a brief overview of the raster sampled prior to the sit-and-stare observations analysed in the following sub-section. In the top row of Fig.~\ref{Loop_Rast}, we plot the intensity of the \ion{Si}{IV} $1394$ \AA\ line at three different positions within the spectral profile. The sunspot brightening within which all super-sonic downflows are detected here manifests as an arc-like structure in the rest wavelength at this time (top left panel), however, at $31$ km s$^{-1}$ (corresponding to approximately $0.15$ \AA\ from the line core) only the northern and southern-most regions of the arc-like structure are evident. Further out into the line at approximately $64$ km s$^{-1}$ (around $0.3$ \AA\ from the line core), only the southern region of the brightening structure has a signature indicating that no super-sonic downflows are present at the northern end of the brightening. In the middle row of Fig.~\ref{Loop_Rast} we plot a spectral representation of this from the \ion{Si}{IV} $1394$ \AA\ line with the intensities taken at the location of the crosses in the upper panels. The super-sonic downflow is evidenced by the secondary peak in intensity to the right of the location of the second vertical dot-dashed line in the right-hand panel. The blue dashed line over-laid plots the Gaussian fit returned for this pixel. The dashed profiles plot the average umbral profile sampled from the blue boxes. The bottom row plots the same as the middle row but for the \ion{O}{IV} $1401$ \AA\ line. Note, the secondary peak in the \ion{O}{IV} $1401$ \AA\ line is not large enough to be returned automatically by the algorithm used here.

\subsection{Sit-And-Stare Sequence}

To continue our analysis, we conducted a more detailed investigation of the signatures of super-sonic downflows within the sunspot in a sit-and-stare observation sampled on the $1$st April $2016$. In Fig.~\ref{Overview_SJI} we plot the FOV at the beginning of the observation, including the SDO/HMI magnetogram and continuum (top left and middle panels), the IRIS chromospheric \ion{Mg}{II} $2796$ \AA\ and transition region \ion{Si}{IV} $1400$ \AA\ filters (top right and bottom left panels, respectively), and the SDO/AIA $304$ \AA\ and $171$ \AA\ channels (bottom middle and right panels). Each panel plots the nearest frame for that channel to $16$:$23$:$28$ UT. The blue contours over-laid on Fig.~\ref{Overview_SJI} map the brightest regions within the sunspot in the $1400$ \AA\ channel, corresponding to the bright arc-like structure plotted in the top left panel of Fig.~\ref{Loop_Rast}. The blue contours on the SDO/HMI continuum panel confirm that this arc-like brightening stretches across the well-developed northern lightbridge. No evidence of this brightening is apparent in the cooler \ion{Mg}{II} $2796$ \AA\ filter. 

\begin{figure*}
\includegraphics[width=0.98\textwidth]{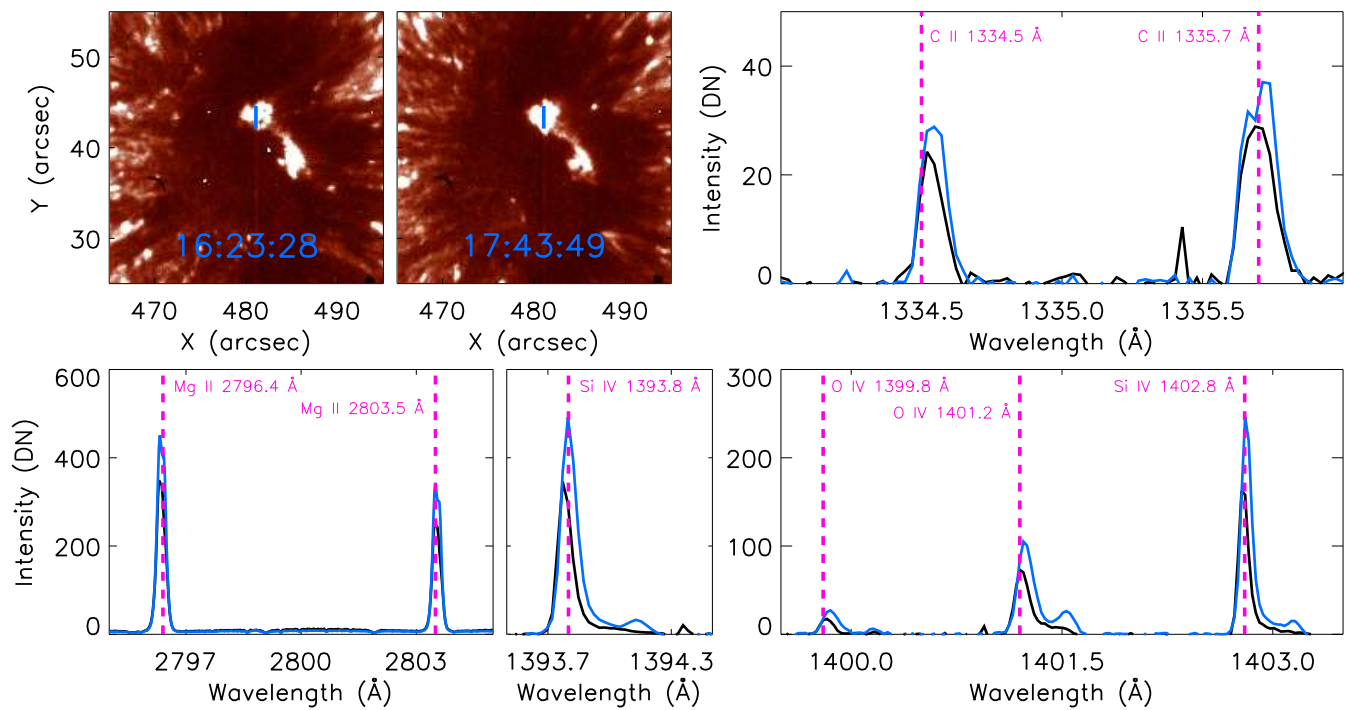}
\caption{Evolution of the SJI and spectral data through time. The first and final SJI frames (top left and top middle panels) show that the transition region brightening remained relatively stable with only a small increase in the size of the northern brightening. The blue vertical lines indicate the slits over which the spectra were averaged in the remaining panels. Spectra from the second and second-to-last raster steps (black and blue lines, respectively) display an increase in intensity in all pass-bands during the super-sonic downflow, however, only the \ion{O}{IV} and \ion{Si}{IV} lines possess the second peak in intensity in the red wing indicative of super-sonic downflows. The vertical dashed pink lines indicate the rest wavelengths of the lines of interest.}
\label{Sp_diff}
\end{figure*}

As can be seen in Fig.~\ref{Overview_SJI}, the IRIS slit passed directly through the northern end of the arc-like brightening during the sit-and-stare sequence, away from the location where super-sonic downflows were detected in the preceding raster (see Fig.~\ref{Loop_Rast}). In Fig.~\ref{Sp_time}, we plot spectral-time plots for both the \ion{Si}{IV} $1394$ \AA\ and the \ion{O}{IV} $1401$ \AA\ lines for the length of the sit-and-stare observation for a pixel within the brightening. Initially, no super-sonic downflow is apparent in either line, agreeing with the spectra sampled for this location during the raster observation (Fig.~\ref{Loop_Rast}); however, a downflow begins to develop with an initial velocity of $\approx72$ km s$^{-1}$ in both the \ion{Si}{IV} $1394$ \AA\ and \ion{O}{IV} $1401$ \AA\ lines after around minute $60$ (corresponding to approximately $17$:$25$ UT). The downflow then accelerates at approximately $5$ m s$^{-2}$ to a velocity of $\approx80$ km s$^{-1}$ over the course of the remaining $\sim20$ minutes, indicated by the dashed blue line. 

In the top left and top middle panels of Fig.~\ref{Sp_diff}, we plot the sunspot as sampled by the IRIS SJI $1400$ \AA\ filter in both the first and the last frames. The arc-like brightening is relatively stable during this time, with only a small increase in brightness and size detected at the north end over the course of the $80$ minutes. Plotting lightcurves of these increases in size and brightness (not shown here) demostrates that they occur relatively consistently throughout the observations and are continuously modulated by the typical $3$-minute sunspot oscillations. Additionally, the ratio of the \ion{Si}{IV} $1394$ \AA\ and $1403$ \AA\ lines within the arc-like brightening throughout the time-series remains close to $2$ implying no changes in the opacity of the lines occurs through time (\citealt{Mathioudakis99}). Overall, no large-scale morphological changes are evident in the arc-like brightening during this time despite the on-set of the super-sonic downflow.

In the other four panels of Fig.~\ref{Sp_diff}, the spectra from the second and second-to-last slit exposures (black and blue lines, respectively) are plotted. For the \ion{C}{II} and \ion{Mg}{II} lines (top right and bottom left panels), increases in intensity of around $10$-$30$ \% are detected between the blue and black profiles, however, no secondary peak associated with super-sonic downflows are evident. This is in agreement with the lack of foot-point signature in the chromospheric \ion{Mg}{II} $2796$ \AA\ SJI data (see top right panel of Fig.~\ref{Overview_SJI}). For the \ion{Si}{IV} and \ion{O}{IV} lines (bottom middle and bottom left panels), no peaks associated with super-sonic downflows are present in the black spectra, however, such peaks are present in the blue spectra with approximate velocities of $80$ km s$^{-1}$. No signature of this downflow was evident in the raster scan starting around $100$ minutes following the end of the sit-and-stare observation.

Interestingly, the arc-like brightening in the transition region coincides spatially with the foot-points of two apparently distinct coronal fan loops in the $171$ \AA\ filter (see Fig.~\ref{Overview_SJI}). A clear separation appears to be present between the two distinct coronal fan loops, with one being anchored in the northern section of the umbra and one being anchored in the southern section of the umbra. A clear gap between these two structures extends to the edge of the FOV (the darker horizontal line at $y\approx42$\arcsec) and beyond. Although it is difficult to track them along their entire lengths, it is possible that the other ends of these two loop systems are rooted in completely separate locations within the Active Region with the northern fan loops appearing to be rooted in the trailing plage region and the southern fan loops potentially rooted in the opposite polarity structures leading the sunspot. Future analysis using Non-Linear Force Free magnetic field extrapolations could better map the connectivity of the Active Region and could, therefore, provide more evidence about where these loops are anchored. Such research could allow us to better examine the potential causes of these downflows.

\section{Discussion and Conclusions}
	\label{Conclusions}

In Table~\ref{Tab1}, the basic properties of the $16$ datasets analysed here were presented. Overall, $13$ of the $16$ datasets included evidence of super-sonic downflows in the \ion{Si}{IV} $1394$ \AA\  line (a similar ratio to that found by \citealt{Samanta18}), with a peak velocity of $112$ km s$^{-1}$ being found on the $29$th March $2016$. In Fig.~\ref{Spect_rast}, the transition region intensity estimated by the rest wavelength of the \ion{Si}{IV} $1394$ \AA\ line was plotted and contours of downflow velocities with magnitudes of $50$ km s$^{-1}$ (aqua) and $75$ km s$^{-1}$ (blue) over-laid. Fig.~\ref{Spect_rast_O} plots the equivalent as calculated from the \ion{O}{IV} $1401$ \AA\ line. All super-sonic downflows detected within the sunspot occurred co-spatial to a sustained transition region brightening spanning across the umbra and penumbra. This transition region brightening appeared to map the locations of coronal loop foot-points, as observed in the SDO/AIA $171$ \AA\ channel. The spatial positioning of super-sonic downflows through time was highly variable, with some rasters revealing relatively large-scale downflows across the entire transition region brightening and some rasters revealing only point-like downflows. Additionally, the spatial coverage of dual flows in the \ion{O}{IV} $1401$ \AA\ line was often different to the spatial coverage in the \ion{Si}{IV} $1394$ \AA\ line, with the spatial coverage in the \ion{O}{IV} $1401$ \AA\ line sometimes being double that of the \ion{Si}{IV} $1394$ \AA\ line.

From both Table~\ref{Tab1} and Fig.~\ref{Int_time}, it is immediately evident that the peak velocity of super-sonic downflows within the umbra is variable through time. This result is in agreement with previous temporal analyses of super-sonic downflows within sunspots (\citealt{Kjeldseth88}) and suggests that single measurements of an AR are unable to provide general information about the properties of super-sonic downflows. The average super-sonic downflow velocity in the \ion{Si}{IV} $1394$ \AA\ line, in rasters where such downflows were observed, was calculated to be $86.7$ km s$^{-1}$ (consistent with previous measurements using IRIS data by, for example, \citealt{Straus15, Chitta16, Samanta18}). Interestingly, super-sonic downflows were detected both before and after the formation of lightbridges in the sunspot indicating that lightbridges in the umbra are not an essential requirement for the formation of super-sonic downflows in transition region spectra (as speculated by \citealt{Nicolas82}).

Super-sonic downflows were also detected in the \ion{O}{IV} $1401$ \AA\ line at the same spatial locations as the peak \ion{Si}{IV} $1394$ \AA\ downflow velocity in $10$ of the $13$ rasters (pixel co-ordinates in Table~\ref{Tab1}). For the other three rasters, downflows were apparent in the \ion{O}{IV} $1401$ \AA\ in the surrounding pixels again highlighting the differences in spatial structuring between these two lines. Additionally, one raster displayed evidence of a super-sonic downflow in the \ion{O}{IV} $1401$ \AA\ when no downflow was apparent in the \ion{Si}{IV} $1394$ \AA\ line. As has been found in previous analyses  (see, for example, \citealt{Samanta18}), the downflow velocities in the \ion{O}{IV} $1401$ \AA\ line were similar to those inferred from the \ion{Si}{IV} $1394$ \AA\ line. One example of a super-sonic downflow in the \ion{C}{II} $1334$ \AA\ line was detected at these locations, with the estimated downflow velocity of this line being $15$ km s$^{-1}$ larger than the downflow inferred from the \ion{Si}{IV} $1394$ \AA\ line at the same position. No signature was detected in the \ion{Mg}{II} $2796$ \AA\ lines at any of these locations. Overall, our results could be interpreted as evidence for the presence of multi-thermal, multi-threaded loops (see, for example, \citealt{Dere82, Nicolas82}), however, they do not rule out other possibilities. If these spectral lines are forming at different heights in the solar atmosphere, one possible interpretation could be that we are observing the acceleration of material down from the hotter regions (\ion{Si}{IV} and \ion{O}{IV} lines) to the cooler chromospheric regions (\ion{C}{II} lines). This would have to be examined further using a range of techniques including RH modelling (\citealt{Uitenbroek01}).

Using an $82$-minute sit-and-stare dataset, we were able to analyse the shorter-scale evolution of super-sonic downflows within the bright sunspot structure. The IRIS slit passed directly through the northern end of the \ion{Si}{IV} transition region brightening that hosted the super-sonic downflows analysed here, which had no signature in the \ion{Mg}{II} filter, at a position where no super-sonic downflows were detected in the previous raster (as shown in Fig.~\ref{Loop_Rast}). Initially, no downflow was apparent in the \ion{Si}{IV} $1394$ \AA\ spectra, however, after around $60$ minutes a dual flow developed with an initial velocity of $73$ km s$^{-1}$. This downflow then accelerated at a rate of $5$ km s$^{-2}$ over the course of the next $20$ minutes to a velocity of around $80$ km s$^{-1}$, similar to the acceleration rate measured at the foot-point of a coronal loop (\citealt{Chitta16}). In the \ion{O}{IV} $1401$ \AA\ spectra, the downflow displayed similar behaviour during the same time period. It is unclear whether this downflow becomes stable at some point (analogous to the downflow studied by \citealt{Straus15}) due to the end of the sit-and-stare observation. It should be noted that no evidence of this downflow is present in the raster following the sit-and-stare indicating that it has a lifetime of only a few hours at most.

Now, we try to place our results in the context of current understanding of the transition region dynamics of sunspots. Although the intensity of the line core is often larger than the background intensity during these downflows, the specific cases where the line core intensity is reduced (see Fig.~\ref{Loop_Rast}), the prevalence of these signatures in the umbra, and the presence of the secondary emission peak in the red wing appears to differentiate the downflows studied here from bright dots (\citealt{Tian14bd}). The sustained nature of the super-sonic downflow in the sit-and-stare observation, the relatively low downflow velocity, and the lack of signature in the \ion{C}{II} $1335$ \AA\ and \ion{Mg}{II} $2796$ \AA\ lines for all but one of the detected dual flows, also makes it unlikely that these downflows are a result of bursty events such as coronal rain (\citealt{Kleint14}). The super-sonic downflows reported here appear to be more similar in velocity and spectral signature to the lower velocity downflows reported in the IRIS literature by several authors (e.g., \citealt{Tian14, Straus15, Chitta16, Samanta18}). It has been suggested that such dual flow signatures in the transition region spectra could be caused by shocks from downflowing material (\citealt{Brynildsen01}), potentially, due to siphon flows (\citealt{Cargill80}). If this is the case then our results imply that the conditions required for siphon flows must be common within ARs throughout their lifetimes.

\section{Summary}
	\label{Summary}

Downflows of over $50$ km s$^{-1}$ were present within the lead sunspot of AR $12526$ in the \ion{Si}{IV} $1394$ \AA\ line for $13$ of the $16$ raster datasets studied here (see Table~\ref{Tab1}). Super-sonic downflows were present in the \ion{O}{IV} $1401$ \AA\ line for $14$ of these datasets. All such downflows occurred co-spatial to a sustained transition region brightening, mapping the foot-points of coronal loops detectable in the SDO/AIA $171$ \AA\ channel, stretching over the umbra and the penumbra. The spatial extents of the downflows in the \ion{Si}{IV} $1394$ \AA\ and \ion{O}{IV} $1401$ \AA\ lines was often very different (Fig.~\ref{Spect_rast} and Fig.~\ref{Spect_rast_O}). The three rasters which contained no evidence of super-sonic downflows within the \ion{Si}{IV} $1394$ \AA\ line occurred following the formation of two lightbridges within the umbra. Only one example of super-sonic downflows was detected in the \ion{C}{II} $1335$ \AA\ line with a velocity $15$ km s$^{-1}$ larger than the corresponding \ion{Si}{IV} $1394$ \AA\ velocity. No downflows were detected in the \ion{Mg}{II} $2796$ \AA\ line.

During the sit-and-stare sequence sampled on the $1$st April $2016$, no super-sonic downflow was initially present (Fig.~\ref{Loop_Rast} and Fig.~\ref{Sp_time}). However, a dual flow did form around $60$ minutes after the start of the observation with an initial velocity of $73$ km s$^{-1}$ in the \ion{Si}{IV} $1394$ \AA\ line. The velocity of this downflow then increased at around $5$ m s$^{-2}$ over the course of the next $20$ minutes to give a final velocity of close to $80$ km s$^{-1}$ (top panel of Fig.~\ref{Sp_time}). In the \ion{O}{IV} $1401$ \AA\ line, the velocity accelerated at a similar rate during the same time. No signature of this downflow was present in either the \ion{C}{II} $1335$ \AA\ or \ion{Mg}{II} $2796$ \AA\ lines (Fig.~\ref{Sp_diff}). Aligning the SJI images to SDO/AIA and SDO/HMI data allowed us to infer that the brightening over which all super-sonic downflows were detected in the rasters appeared to stretch over the northern lightbridge at this time. Additionally, two distinct sets of coronal fan loops were evident with one set appearing to be rooted to the north of this lightbridge and one set appearing to be rooted south of this lightbridge (Fig.~\ref{Overview_SJI}).

Overall, our results indicate that super-sonic downflows are highly irregular and intermittent over the course of both minutes and hours (agreeing with previous research by \citealt{Kjeldseth88}). The spatial locations and velocities of these downflows, which are most easily detected in the red wing of the \ion{Si}{IV} $1394$ \AA\ and \ion{O}{IV} $1401$ \AA\ lines, are often completely different from one raster to the next, and from one spectral window to another. Our analysis also suggests that super-sonic downflows occur preferentially around bright regions at the \ion{Si}{IV} $1394$ \AA\ rest wavelength corresponding to the locations of foot-points of coronal loops. Finally, the intermittency of super-sonic downflows implies that more sunspots than previously thought could host such events (i.e, more than the $\sim80$ \% reported by \citealt{Samanta18}).

\begin{acknowledgements}
We thank the Science and Technology Facilities Council (STFC) for the support received to conduct this research through grant number: ST/P000304/1. SKP is grateful to the FWO Vlaanderen for a senior postdoctoral fellowship. IRIS is a NASA small explorer mission developed and operated by LMSAL with mission operations executed at NASA Ames Research Center and major contributions to downlink communications funded by ESA and the Norwegian Space Centre. SDO/HMI and SDO/AIA data are courtesy of NASA/SDO and the HMI and AIA science teams.
\end{acknowledgements}

\bibliographystyle{aa}
\bibliography{Umbral_Loop}

\begin{thebibliography}{34}
\expandafter\ifx\csname natexlab\endcsname\relax\def\natexlab#1{#1}\fi

\bibitem[{{Beckers} \& {Tallant}(1969)}]{Beckers69}
{Beckers}, J.~M. \& {Tallant}, P.~E. 1969, \solphys, 7, 351

\bibitem[{{Bose} {et~al.}(2019){Bose}, {Henriques}, {Rouppe van der Voort}, \&
  {Pereira}}]{Bose19}
{Bose}, S., {Henriques}, V. M.~J., {Rouppe van der Voort}, L., \& {Pereira}, T.
  M.~D. 2019, \aap, 627, A46

\bibitem[{{Brynildsen} {et~al.}(2001){Brynildsen}, {Maltby}, {Kjeldseth-Moe},
  \& {Wilhelm}}]{Brynildsen01}
{Brynildsen}, N., {Maltby}, P., {Kjeldseth-Moe}, O., \& {Wilhelm}, K. 2001,
  \apjl, 552, L77

\bibitem[{{Brynildsen} {et~al.}(2004){Brynildsen}, {Maltby}, {Kjeldseth-Moe},
  \& {Wilhelm}}]{Brynildsen04}
{Brynildsen}, N., {Maltby}, P., {Kjeldseth-Moe}, O., \& {Wilhelm}, K. 2004,
  \apj, 612, 1193

\bibitem[{{Cargill} \& {Priest}(1980)}]{Cargill80}
{Cargill}, P.~J. \& {Priest}, E.~R. 1980, \solphys, 65, 251

\bibitem[{{Chitta} {et~al.}(2016){Chitta}, {Peter}, \& {Young}}]{Chitta16}
{Chitta}, L.~P., {Peter}, H., \& {Young}, P.~R. 2016, \aap, 587, A20

\bibitem[{{De Pontieu} {et~al.}(2014){De Pontieu}, {Title}, {Lemen}, {Kushner},
  {Akin}, {Allard}, {Berger}, {Boerner}, {Cheung}, {Chou}, {Drake}, {Duncan},
  {Freeland}, {Heyman}, {Hoffman}, {Hurlburt}, {Lindgren}, {Mathur}, {Rehse},
  {Sabolish}, {Seguin}, {Schrijver}, {Tarbell}, {W{\"u}lser}, {Wolfson},
  {Yanari}, {Mudge}, {Nguyen-Phuc}, {Timmons}, {van Bezooijen}, {Weingrod},
  {Brookner}, {Butcher}, {Dougherty}, {Eder}, {Knagenhjelm}, {Larsen},
  {Mansir}, {Phan}, {Boyle}, {Cheimets}, {DeLuca}, {Golub}, {Gates}, {Hertz},
  {McKillop}, {Park}, {Perry}, {Podgorski}, {Reeves}, {Saar}, {Testa}, {Tian},
  {Weber}, {Dunn}, {Eccles}, {Jaeggli}, {Kankelborg}, {Mashburn}, {Pust},
  {Springer}, {Carvalho}, {Kleint}, {Marmie}, {Mazmanian}, {Pereira}, {Sawyer},
  {Strong}, {Worden}, {Carlsson}, {Hansteen}, {Leenaarts}, {Wiesmann},
  {Aloise}, {Chu}, {Bush}, {Scherrer}, {Brekke}, {Martinez-Sykora}, {Lites},
  {McIntosh}, {Uitenbroek}, {Okamoto}, {Gummin}, {Auker}, {Jerram}, {Pool}, \&
  {Waltham}}]{dePontieu14}
{De Pontieu}, B., {Title}, A.~M., {Lemen}, J.~R., {et~al.} 2014, \solphys, 289,
  2733

\bibitem[{{Dere}(1982)}]{Dere82}
{Dere}, K.~P. 1982, \solphys, 77, 77

\bibitem[{{Fludra}(2001)}]{Fludra01}
{Fludra}, A. 2001, \aap, 368, 639

\bibitem[{{Hale}(1908)}]{Hale08}
{Hale}, G.~E. 1908, \apj, 28, 315

\bibitem[{{Henriques} {et~al.}(2017){Henriques}, {Mathioudakis},
  {Socas-Navarro}, \& {de la Cruz Rodr{\'\i}guez}}]{Henriques17}
{Henriques}, V.~M.~J., {Mathioudakis}, M., {Socas-Navarro}, H., \& {de la Cruz
  Rodr{\'\i}guez}, J. 2017, \apj, 845, 102

\bibitem[{{Jess} {et~al.}(2012){Jess}, {De Moortel}, {Mathioudakis},
  {Christian}, {Reardon}, {Keys}, \& {Keenan}}]{Jess12}
{Jess}, D.~B., {De Moortel}, I., {Mathioudakis}, M., {et~al.} 2012, \apj, 757,
  160

\bibitem[{{Khomenko} \& {Collados}(2015)}]{Khomenko15}
{Khomenko}, E. \& {Collados}, M. 2015, Living Reviews in Solar Physics, 12, 6

\bibitem[{{Kjeldseth-Moe} {et~al.}(1988){Kjeldseth-Moe}, {Brynildsen},
  {Brekke}, {Engvold}, {Maltby}, {Bartoe}, {Brueckner}, {Cook}, {Dere}, \&
  {Socker}}]{Kjeldseth88}
{Kjeldseth-Moe}, O., {Brynildsen}, N., {Brekke}, P., {et~al.} 1988, \apj, 334,
  1066

\bibitem[{{Kleint} {et~al.}(2014){Kleint}, {Antolin}, {Tian}, {Judge}, {Testa},
  {De Pontieu}, {Mart{\'\i}nez-Sykora}, {Reeves}, {Wuelser}, {McKillop},
  {Saar}, {Carlsson}, {Boerner}, {Hurlburt}, {Lemen}, {Tarbell}, {Title},
  {Golub}, {Hansteen}, {Jaeggli}, \& {Kankelborg}}]{Kleint14}
{Kleint}, L., {Antolin}, P., {Tian}, H., {et~al.} 2014, \apjl, 789, L42

\bibitem[{{Lagg} {et~al.}(2007){Lagg}, {Woch}, {Solanki}, \& {Krupp}}]{Lagg07}
{Lagg}, A., {Woch}, J., {Solanki}, S.~K., \& {Krupp}, N. 2007, \aap, 462, 1147

\bibitem[{{Lemen} {et~al.}(2012){Lemen}, {Title}, {Akin}, {Boerner}, {Chou},
  {Drake}, {Duncan}, {Edwards}, {Friedlaender}, {Heyman}, {Hurlburt}, {Katz},
  {Kushner}, {Levay}, {Lindgren}, {Mathur}, {McFeaters}, {Mitchell}, {Rehse},
  {Schrijver}, {Springer}, {Stern}, {Tarbell}, {Wuelser}, {Wolfson}, {Yanari},
  {Bookbinder}, {Cheimets}, {Caldwell}, {Deluca}, {Gates}, {Golub}, {Park},
  {Podgorski}, {Bush}, {Scherrer}, {Gummin}, {Smith}, {Auker}, {Jerram},
  {Pool}, {Soufli}, {Windt}, {Beardsley}, {Clapp}, {Lang}, \&
  {Waltham}}]{Lemen12}
{Lemen}, J.~R., {Title}, A.~M., {Akin}, D.~J., {et~al.} 2012, \solphys, 275, 17

\bibitem[{{Mathioudakis} {et~al.}(1999){Mathioudakis}, {McKenny}, {Keenan},
  {Williams}, \& {Phillips}}]{Mathioudakis99}
{Mathioudakis}, M., {McKenny}, J., {Keenan}, F.~P., {Williams}, D.~R., \&
  {Phillips}, K.~J.~H. 1999, \aap, 351, L23

\bibitem[{{Nelson} {et~al.}(2017){Nelson}, {Henriques}, {Mathioudakis}, \&
  {Keenan}}]{Nelson17}
{Nelson}, C.~J., {Henriques}, V.~M.~J., {Mathioudakis}, M., \& {Keenan}, F.~P.
  2017, \aap, 605, A14

\bibitem[{{Nicolas} {et~al.}(1982){Nicolas}, {Bartoe}, {Brueckner}, \&
  {Kjeldseth-Moe}}]{Nicolas82}
{Nicolas}, K.~R., {Bartoe}, J. D.~F., {Brueckner}, G.~E., \& {Kjeldseth-Moe},
  O. 1982, \solphys, 81, 253

\bibitem[{{Priest}(1984)}]{Priest84}
{Priest}, E.~R. 1984, {Solar magneto-hydrodynamics}

\bibitem[{{Rouppe van der Voort} \& {de la Cruz
  Rodr{\'\i}guez}(2013)}]{Rouppe13}
{Rouppe van der Voort}, L. \& {de la Cruz Rodr{\'\i}guez}, J. 2013, \apj, 776,
  56

\bibitem[{{Rouppe van der Voort} {et~al.}(2003){Rouppe van der Voort},
  {Rutten}, {S{\"u}tterlin}, {Sloover}, \& {Krijger}}]{Rouppe03}
{Rouppe van der Voort}, L.~H.~M., {Rutten}, R.~J., {S{\"u}tterlin}, P.,
  {Sloover}, P.~J., \& {Krijger}, J.~M. 2003, \aap, 403, 277

\bibitem[{{Samanta} {et~al.}(2018){Samanta}, {Tian}, \& {Prasad
  Choudhary}}]{Samanta18}
{Samanta}, T., {Tian}, H., \& {Prasad Choudhary}, D. 2018, \apj, 859, 158

\bibitem[{{Scherrer} {et~al.}(2012){Scherrer}, {Schou}, {Bush}, {Kosovichev},
  {Bogart}, {Hoeksema}, {Liu}, {Duvall}, {Zhao}, {Title}, {Schrijver},
  {Tarbell}, \& {Tomczyk}}]{Scherrer12}
{Scherrer}, P.~H., {Schou}, J., {Bush}, R.~I., {et~al.} 2012, \solphys, 275,
  207

\bibitem[{{Socas-Navarro} {et~al.}(2015){Socas-Navarro}, {de la Cruz
  Rodr{\'\i}guez}, {Asensio Ramos}, {Trujillo Bueno}, \& {Ruiz Cobo}}]{Socas15}
{Socas-Navarro}, H., {de la Cruz Rodr{\'\i}guez}, J., {Asensio Ramos}, A.,
  {Trujillo Bueno}, J., \& {Ruiz Cobo}, B. 2015, \aap, 577, A7

\bibitem[{{Solanki}(2003)}]{Solanki03}
{Solanki}, S.~K. 2003, \aapr, 11, 153

\bibitem[{{Straus} {et~al.}(2015){Straus}, {Fleck}, \& {Andretta}}]{Straus15}
{Straus}, T., {Fleck}, B., \& {Andretta}, V. 2015, \aap, 582, A116

\bibitem[{{Sych} \& {Nakariakov}(2014)}]{Sych14}
{Sych}, R. \& {Nakariakov}, V.~M. 2014, \aap, 569, A72

\bibitem[{{Tian} {et~al.}(2014{\natexlab{a}}){Tian}, {DeLuca}, {Reeves},
  {McKillop}, {De Pontieu}, {Mart{\'\i}nez-Sykora}, {Carlsson}, {Hansteen},
  {Kleint}, {Cheung}, {Golub}, {Saar}, {Testa}, {Weber}, {Lemen}, {Title},
  {Boerner}, {Hurlburt}, {Tarbell}, {Wuelser}, {Kankelborg}, {Jaeggli}, \&
  {McIntosh}}]{Tian14}
{Tian}, H., {DeLuca}, E., {Reeves}, K.~K., {et~al.} 2014{\natexlab{a}}, \apj,
  786, 137

\bibitem[{{Tian} {et~al.}(2014{\natexlab{b}}){Tian}, {Kleint}, {Peter},
  {Weber}, {Testa}, {DeLuca}, {Golub}, \& {Schanche}}]{Tian14bd}
{Tian}, H., {Kleint}, L., {Peter}, H., {et~al.} 2014{\natexlab{b}}, \apjl, 790,
  L29

\bibitem[{{Tian} {et~al.}(2018){Tian}, {Samanta}, \& {Zhang}}]{Tian18}
{Tian}, H., {Samanta}, T., \& {Zhang}, J. 2018, Geoscience Letters, 5, 4

\bibitem[{{Uitenbroek}(2001)}]{Uitenbroek01}
{Uitenbroek}, H. 2001, \apj, 557, 389

\bibitem[{{Wittmann}(1969)}]{Wittmann69}
{Wittmann}, A. 1969, \solphys, 7, 366

\end{thebibliography}

\end{document}